\documentclass[submission, Proceedings]{SciPost}

\hypersetup{
    colorlinks,
    linkcolor={red!50!black},
    citecolor={blue!50!black},
    urlcolor={blue!80!black}
}

\usepackage{float}
\usepackage[bitstream-charter]{mathdesign}
\DeclareSymbolFont{usualmathcal}{OMS}{cmsy}{m}{n}
\DeclareSymbolFontAlphabet{\mathcal}{usualmathcal}

\begin{document}

\begin{center}{\Large \textbf{
Transverse Single Spin Asymmetries of Heavy Flavor Electrons and Charged Pions in 200 GeV $p+p^{\uparrow}$ Collisions at Midrapidity
}}\end{center}

\begin{center}
Dillon S. Fitzgerald\textsuperscript{1$\dag\star$}
\end{center}

\begin{center}
{\bf 1} University of Michigan, Ann Arbor, USA
\\
$\dag$ On behalf of the PHENIX collaboration \\
$\star$ dillfitz@umich.edu 
\end{center}

\begin{center}
\today
\end{center}


\definecolor{palegray}{gray}{0.95}
\begin{center}
\colorbox{palegray}{
  \begin{tabular}{rr}
  \begin{minipage}{0.1\textwidth}
    \includegraphics[width=22mm]{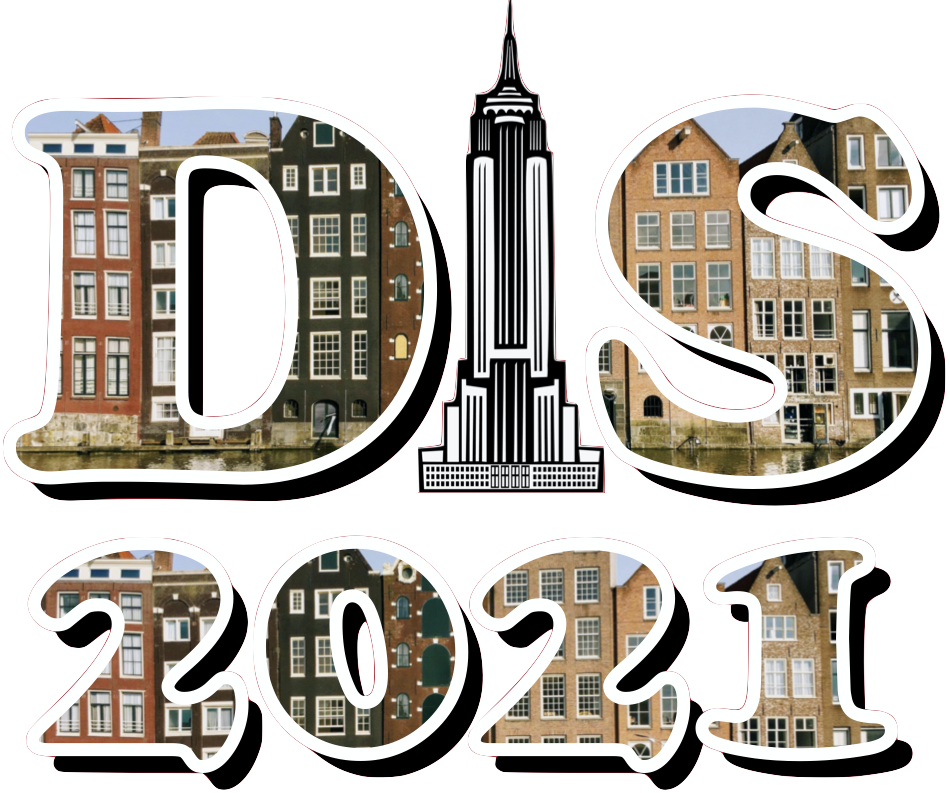}
  \end{minipage}
  &
  \begin{minipage}{0.75\textwidth}
    \begin{center}
    {\it Proceedings for the XXVIII International Workshop\\ on Deep-Inelastic Scattering and
Related Subjects,}\\
    {\it Stony Brook University, New York, USA, 12-16 April 2021} \\
    \doi{10.21468/SciPostPhysProc.?}\\
    \end{center}
  \end{minipage}
\end{tabular}
}
\end{center}

\section*{Abstract}
{\bf
    Transverse single spin asymmetries of particles produced in $p+p^{\uparrow}$ collisions provide insight on the partonic spin and momentum structure of hadrons; heavy flavor electrons provide access to initial state spin-momentum correlations of gluons in the proton, while charged pions provide access to initial and final state transverse spin effects. Charged particles are measured at midrapidity at PHENIX using the silicon vertex detector and central arm spectrometer, made of an electromagnetic calorimeter, a ring-imaging Cherenkov detector, and drift and pad chambers. Recent results for both electron and charged pion measurements from the 2015 running period will be presented.
}

\vspace{10pt}
\noindent\rule{\textwidth}{1pt}
\tableofcontents\thispagestyle{fancy}
\noindent\rule{\textwidth}{1pt}
\vspace{10pt}
\section{Introduction}
\label{sec:intro}
Our understanding of proton structure has evolved greatly over the past few decades --  from a naive picture where the valence quarks carry all of the proton's spin, to a much more complex picture involving spin and orbital angular momentum of quarks and gluons. The complex structure of hadrons gives rise to emergent properties of bound states such as spin-momentum and spin-spin correlations within hadronic systems, similar to spin-orbit and spin-spin couplings in atomic systems. These correlations are manifest in experimentally observable ways, such as azimuthal modulations of particle yields in collisions involving polarized hadrons. One such observable, the transverse single-spin asymmetry (TSSA), involves collisions with an unpolarized beam particle and a transversely polarized hadron in the initial state (e.g. $p+p^{\uparrow}$ at the Relativistic Heavy Ion Collider (RHIC)), and is defined as follows, 
\begin{equation}
A_{N} (\phi) = \frac{\sigma^{\uparrow}(\phi) - \sigma^{\downarrow} (\phi)}{\sigma^{\uparrow}(\phi) + \sigma^{\downarrow} (\phi)} = A_{N} \cos\phi.
\end{equation}\label{eqn:AN_phi}
$\sigma^{\uparrow,\downarrow}(\phi)$ are cross sections measured as a function of $\phi$ for collision systems with $p^{\uparrow}$ and $p^{\downarrow}$ initial state protons respectively~\cite{muons}. Perturbative Quatnum Chromodynamics (pQCD) calculations predict negligible contributions to TSSAs from partonic hard scattering effects (<1\%)~\cite{kane}. However, large TSSAs have been measured as a function of $x_{F} = 2p_{z}/\sqrt{s}$ and $p_{T}$ of produced particles, even persisting at $\sqrt{s} = 200$ GeV and $p_{T}$ $\approx 7$ GeV/c, well into the regime where pQCD applies\cite{largeTSSA,KIT}. This implies the existence of nonperturbative spin-momentum and spin-spin correlations within hadrons, as contributions to large TSSAs must be coming from the nonperturbative elements of factorized cross sections. 

Two theoretical frameworks have emerged that have been successful in describing the large observed TSSAs, one with transverse momentum dependent (TMD) nonperturbative functions that depend explicitly on the transverse momentum of partons within the hadron, and one with higher twist effects, where higher order power suppressed terms in the collinear factorization expansion must be considered. Both have the notable feature that one must look beyond standard twist-2 collinear factorization -- TMD functions require access to both a hard and soft scale with sufficient scale separation (e.g. $Q>>k_{T}$), while the higher twist framework preserves the collinear factorization scheme, allowing for access from observables sensitive to only the hard scale, which is the case for the measurements presented in this note. Looking to higher twist leads to terms in the polarized cross section including the convolution of correlation functions dependent on momentum fractions of 2 partons in a given hadron with standard parton distribution functions (PDFs) and fragmentation functions (FFs). This leads to the following proportionality for TSSAs in the collinear twist-3 framework~\cite{forwardjpsi},
\begin{equation}
	\begin{aligned}
	    A_{N} \propto	\sum_{a,b,c}\phi^{(3)}_{a/A} (x_{1},x_{2},\vec{s_{\perp}}) \otimes \phi_{b/B} (x') \otimes \hat{\sigma} \otimes D_{c \rightarrow C}(z) \\ + \sum_{a,b,c} \delta q_{a/A} (x,\vec{s_{\perp}}) \otimes\phi^{(3)}_{b/B} (x'_{1},x'_{2}) \otimes \hat{\sigma '}\otimes D_{c \rightarrow C} (z) \\ + \sum_{a,b,c} \delta q_{a/A} (x,\vec{s_{\perp}}) \otimes \phi_{b/B} (x') \otimes \hat{\sigma ''} \otimes D^{(3)}_{c \rightarrow C} (z_{1},z_{2}). 	   
	\end{aligned}\label{eqn:AN_twist3}            
\end{equation}
Here $\delta q_{a/A}$ is the transversity distribution of parton $\it{a}$ in polarized hadron $\it{A}$, and any term with a superscript $(3)$ is a twist-3 correlator, also known as a parton correlation function (PCF). Unification of these two theoretical frameworks has been demonstrated, with twist 3 correlators related to moments in transverse momentum of the corresponding parton TMD~\cite{pitonyak}. Hence one can learn something interesting about different spin-momentum and spin-spin correlations from each term in Equation~\ref{eqn:AN_twist3}. The first term is related to the Sivers TMD, while the second is related to transversity $\otimes$ Boer-Mulders TMD and the third to transversity $\otimes$ Collins TMD FF ~\cite{twist3_TMD_types}. By measuring $A_{N}$ for particular final state particles, different elements of Equation~\ref{eqn:AN_twist3} can be isolated and constrained. Heavy flavor electron production is dominated by gluon-gluon fusion at $\sqrt{s} = 200$ GeV at midrapidity, and the transversity distribution for gluons $\delta_{g/X} = 0$, providing access to the twist 3 correlator in the polarized proton $\phi^{(3)}_{g/A} (x_{1},x_{2},\vec{s_{\perp}})$ while the two terms that include convolution with transversity distributions are zero~\cite{forwardjpsi}. On the other hand, charged pion production is dominated by quark-gluon scattering in the same kinematic region, providing broader access to the different terms in Equation~\ref{eqn:AN_twist3} and sensitivity to quark flavor.	

\section{Experimental Setup}\label{sec:ana}
Data for the analyses presented here were taken from PHENIX at RHIC. In particular, the measurements presented are from the 2015 running year with $\sqrt{s} = 200$ GeV $p+p^{\uparrow}$ initial state, and inclusive final state particles measured at midrapidity $\mid \eta \mid < 0.35$.

\subsection{Charged Particle Identification at PHENIX}\label{sec:detector}
The PHENIX detector contains two central arm spectrometers with acceptance $\mid \eta \mid < 0.35$, and $\Delta \phi = 0.5$ per arm. Each consist of a drift chamber and pad chambers for tracking and momentum measurements, ring-imaging Cherenkov (RICH) detectors for particle identification, and electromagnetic calorimeters (EMCal) for measuring energy deposition and triggering on charged particles. In addition, for the open heavy flavor electron analysis, the hit pattern in the silicon vertex detector close to the beam line is measured in order to veto conversion electrons. All of these components allow for curation of samples of charge particles at PHENIX, and additional information from the RICH and EMCal subsystems can be used to separate samples of charged hadrons from electrons. The RICH detector has a Cherenkov threshold of $\gamma = 35$, providing $e$, $\pi$ separation from $20$ MeV/c $< p_{T} < 5$ GeV/c. Furthermore, electrons deposit all or most of their energy in the EMCal, while charged hadrons only deposit a fraction of their energy, allowing one to use the $E/p$ distribution of charged tracks to improve electron purity. 
\subsection{Analysis Procedure}\label{sec:procedure}
The TSSA observable $A_{N}$ is calculated with the relative luminosity formula, integrating over the $\phi$ ranges of the east and west spectrometer arm due to the limited detector acceptance, 
\begin{equation}
A_{N} = \frac{1}{\langle \mid \cos \phi \mid \rangle }\frac{1}{P}\frac{N_{L}^{\uparrow} - R N_{L}^{\downarrow}}{N_{L}^{\uparrow} + R N_{L}^{\downarrow}}\label{eqn:AN_lumi}.
\end{equation}
The $\uparrow,\downarrow$ superscripts represent spin up vs spin down bunch crossings, the $L$ subscript represents the the left spectrometer arm, $P$ is the average polarization fraction of the beam, $\langle \mid \cos \phi \mid \rangle$ is a correction factor applied for integrating over $\phi$, and $R = \mathcal{L}^{\uparrow}/\mathcal{L}^{\downarrow}$ is the relative luminosity, defined as the ratio of luminosity for spin up bunch crossings vs spin down bunch crossings. Note that there is an equivalent formula to Equation~\ref{eqn:AN_lumi} for the opposite spectrometer arm that comes with an additional factor of $-1$. As mentioned in Section~\ref{sec:detector}, the central arm spectrometers measure charged particle yields. In addition to this, the luminosity $\mathcal{L}$ is measured by the beam beam counter (BBC), and beam polarization $P$ with the RHIC polarimeters, giving all the necessary ingredients for Equation~\ref{eqn:AN_lumi}. Once $A^{S+B}_{N}$ is calculated for an inclusive sample, it is necessary to quantify the signal to background ratio in each bin as well as the background asymmetries $A_{N}^{B_{i}}$ in order to extract the signal asymmetry $A^{S}_{N}$. For the heavy flavor electron analysis, the backgrounds consist of electrons from other sources ($\pi^{0}, \eta, \gamma, J/\psi, K^{0}_{S}/K^{\pm}$), and charged hadrons (primarily $\pi^{\pm}$), while for the $\pi^{\pm}$ analysis, the backgrounds consist of mostly electrons, with additional contributions from heavier hadrons ($p/\bar{p}, K^{\pm}$). However, these heavier hadrons contribute negligibly in the measured range of the $\pi^{\pm}$ results ($5 $GeV/c$ < p_{T} < 15 $GeV/c). An electron cocktail method is used to estimate background fractions of various electronic decay channels in simulation, and data driven methods can be applied using information recorded in the EMCal and RICH detector subsystems to quantify the $e^{\pm}/h^{\pm}$ fraction. The EMCal and RICH selection requirements can be tuned to balance electron purity vs. selection efficiency. The background asymmetries for the heavy flavor analysis were taken from~\cite{pi0AN, dpAN} for the photonic sources ($\pi^{0}, \eta, \gamma$), from~\cite{hadronAN} for the hadron contamination $h^{\pm}$, and ~\cite{jpsiAN} for the $J/\psi$, while $K^{0}_{S}/K^{\pm}$ was measured to be negligible in the measured $p_{T}$ range, and was not considered in the background correction. The background asymmetry for the $J/\psi$ is a large source of statistical uncertainty for the final heavy flavor electron results. For this reason, a non photonic electron asymmetry was also measured that only takes $(\pi^{0}, \eta, \gamma, h^{\pm})$ into account in the background correction procedure. 

\section{Heavy Flavor Electron TSSA}\label{sec:electrons}
\begin{figure}[H]
    \centering
    \includegraphics[width=0.5\textwidth]{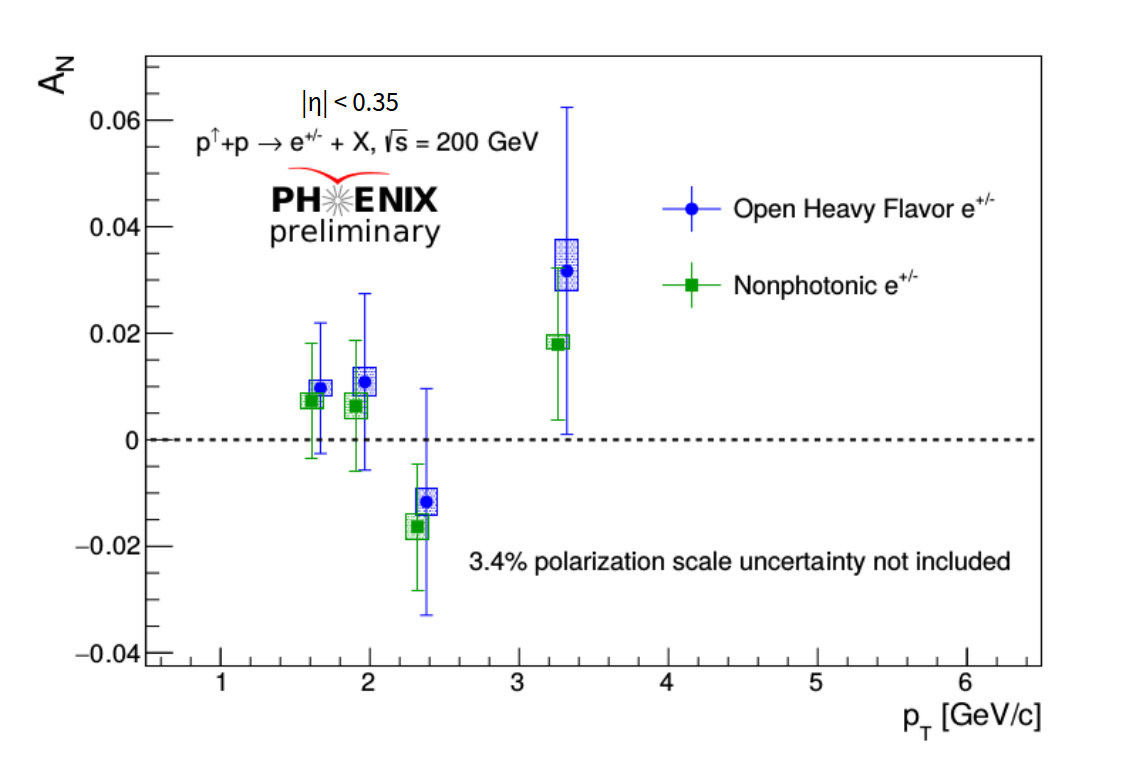}
    \caption{Heavy flavor $e^{\pm}$ $A_{N}(p_{T})$ (blue), nonphotonic $e^{\pm}$ $A_{N}(p_{T})$ (green)}
    \label{fig:AN_chargecombined}
\end{figure}

\begin{figure}[H]
    \centering
    \includegraphics[width=0.99\textwidth]{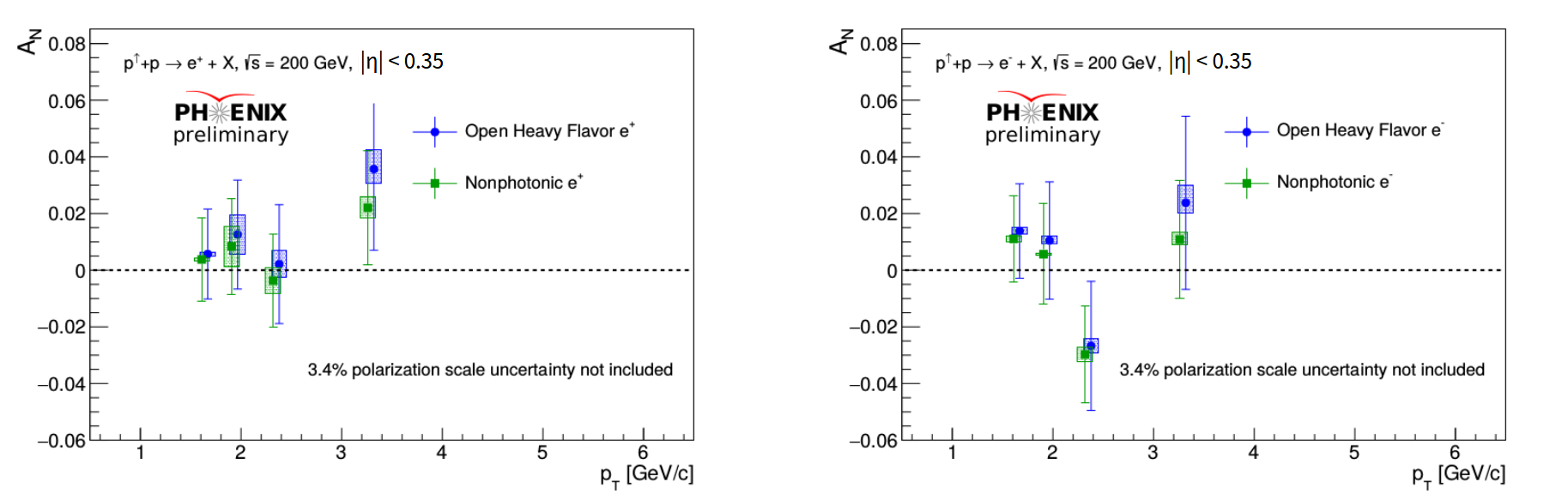}
    \caption{Left panel: heavy flavor $e^{+}$ $A_{N}(p_{T})$ (blue), nonphotonic $e^{+}$ $A_{N}(p_{T})$ (green). Right panel: Heavy flavor $e^{-}$ $A_{N}(p_{T})$ (blue), nonphotonic $e^{-}$ $A_{N}(p_{T})$ (green) }
    \label{fig:AN_chargesep}
\end{figure}

\begin{figure}[H]
    \centering
    \includegraphics[width=0.99\textwidth]{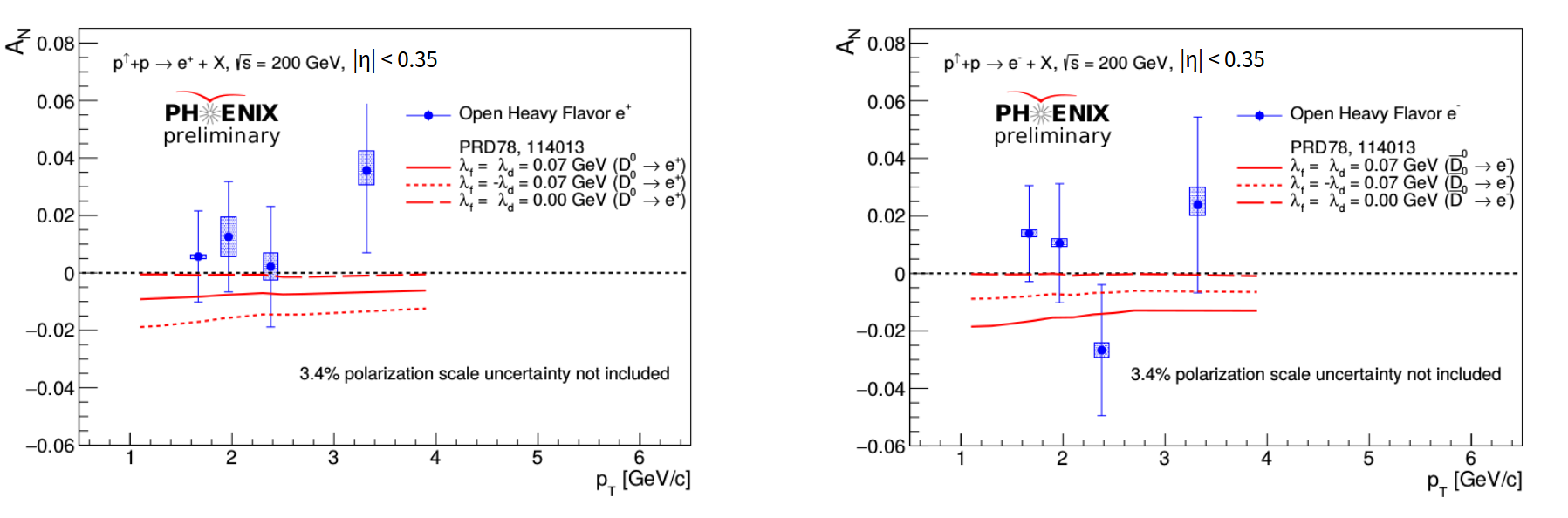}
    \caption{Left panel: heavy flavor $e^{+}$ $A_{N}(p_{T})$ (blue), theoretical predictions of $A_{N}^{D^{0}\rightarrow e^{+}}(p_{T})$ from~\cite{trigluon_twists} + PYTHIA simulations (red). Right panel: heavy flavor $e^{-}$ $A_{N}(p_{T})$ (blue), theoretical predictions of $A_{N}^{\bar{D}^{0}\rightarrow e^{-}}(p_{T})$ from~\cite{trigluon_twists} + PYTHIA simulations (red). }\label{fig:AN_chargesep_theory}
\end{figure}

Heavy flavor production at midrapidity is dominated by the gluon-gluon fusion channel at $\sqrt{s} = 200$ GeV~\cite{trigluon_twists}, receiving only a small contribution from $q \bar{q}$ annihilation in the measured range $1.0$ GeV/c $< p_{T} < 5.0$ GeV/c (note that the RICH provides good $e/\pi$ separation in that range). Open charm in the final state is also dominant at these kinematics. The gluon-gluon fusion channel is of particular interest because the twist 3 trigluon (ggg) correlator is not well constrained from previous measurements, while the Efremov-Teryaev-Qiu-Sterman (qgq) correlator is somewhat constrained from previous measurements. In addition, since the gluon is a massless gauge boson, its transversity distribution is zero, isolating the term involving the trigluon correlator in Equation~\ref{eqn:AN_twist3}. It should be noted that the qgq correlator is used as an input to extract information on the ggg correlator. Figure~\ref{fig:AN_chargecombined} shows the results of the charge combined analysis, with heavy flavor $e^{\pm}$ $A_{N}(p_{T})$ in blue and nonphotonic $e^{\pm}$ $A_{N}(p_{T})$ in green, both are the most statistically precise measurements of this observable at midrapidity. It can be seen that both heavy flavor and nonphotonic $e^{\pm}$ $A_{N}(p_{T})$ results are consistent with zero in the measured range.
\noindent Figure~\ref{fig:AN_chargesep} shows the results of the charge separated analysis, with heavy flavor $e^{+}$ $A_{N}(p_{T})$ in blue and nonphotonic $e^{+}$ $A_{N}(p_{T})$ in green in the left panel and, and similarly for $e^{-}$ $A_{N}(p_{T})$ in the right panel. It can again be observed that both heavy flavor and nonphotonic $e^{+,-}$ $A_{N}(p_{T})$ results are consistent with zero in the measured range.
\noindent Most interestingly, Figure~\ref{fig:AN_chargesep_theory} shows the results of the charge separated analysis again for only the heavy flavor $e^{+,-}$ samples, this time plotted alongside theoretical results from~\cite{trigluon_twists}. $A_{N}^{D^{0},\bar{D}^{0}}(p_{T})$ were calculated at midrapidity in~\cite{trigluon_twists} and PYTHIA simulations were conducted to convert these asymmetries to $A_{N}^{D^{0}\rightarrow e^{+},\bar{D}^{0} \rightarrow e^{-}} (p_{T})$. In particular, the left panel shows $e^{+}$ $A_{N}(p_{T})$ in blue plotted alongside theoretical predictions of $A_{N}^{D^{0}\rightarrow e^{+}}(p_{T})$ from~\cite{trigluon_twists} + PYTHIA simulations for various parameter combinations $(\lambda_{f}, \lambda_{d})$, while the right panel shows $e^{-}$ $A_{N}(p_{T})$ in blue plotted alongside theoretical predictions of $A_{N}^{\bar{D}^{0}\rightarrow e^{-}}(p_{T})$ from~\cite{trigluon_twists} + PYTHIA simulations those same parameter combinations $(\lambda_{f}, \lambda_{d})$. The $\lambda$ parameters correspond to normalization factors of trigluon correlators $\phi^{(3),(f,d)}_{g/A} (x,x,\vec{s_{\perp}})$ at a particular fixed point in x, to unpolarized gluon PDFs $g(x)$, with $\phi^{(3),f}_{g/A} (x,x,\vec{s_{\perp}})$ = $\lambda_{f} g(x)$ and $\phi^{(3),d}_{g/A} (x,x,\vec{s_{\perp}})$ = $\lambda_{d} g(x)$. The theoretical curves are ordered differently for the different charges, as shown in Figure~\ref{fig:AN_chargesep_theory}, allowing for some sensitivity to constrain $\lambda_{f}$ and $\lambda_{d}$ by looking at the charges separately.

\section{Charged Pion TSSA}\label{sec:pions}
\begin{figure}[H]
    \centering
    \includegraphics[width=0.5\textwidth]{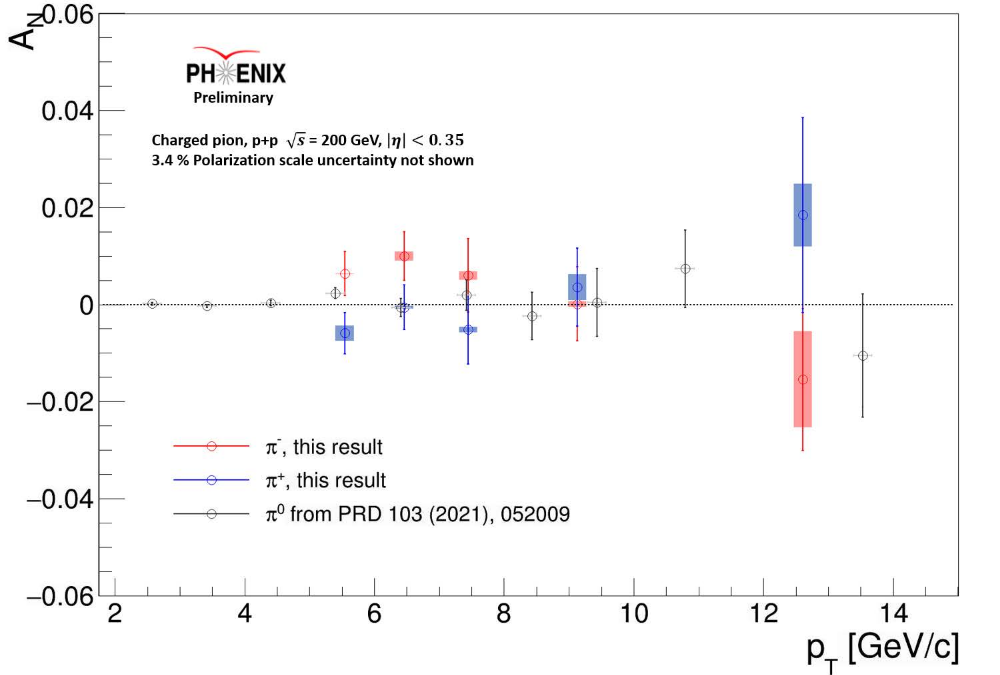}
    \caption{$\pi^{-}$ $A_{N}(p_{T})$ (red), $\pi^{+}$ $A_{N}(p_{T})$ (blue), $\pi^{0}$ $A_{N}(p_{T})$ from~\cite{pi0AN} (gray)}
    \label{fig:AN_chargedpions}
\end{figure}

Charged pion production at midrapidity is dominated by quark-gluon scattering at $\sqrt{s} = 200$ GeV in the measured range $5$ GeV/c $ < p_{T} < 15$ GeV/c, potentially allowing for sensitivity to light quark flavors when looking at $\pi^{\pm}$ separately. In addition, $(\pi^{\pm}, \pi^{0})$ is an isospin triplet; comparing $A_{N}$ for these different particles is a good test for theoretical models. The $\pi^{0}$ $A_{N}(p_{T})$ was previously measured at $\sqrt{s} = 200$ GeV at midrapidity~\cite{pi0AN}. This is compared with new results for $\pi^{\pm}$ $A_{N}(p_{T})$ in Figure~\ref{fig:AN_chargedpions}, with the red points representing $\pi^{-}$ $A_{N}(p_{T})$, blue points representing $\pi^{+}$ $A_{N}(p_{T})$, and gray points representing $\pi^{0}$ $A_{N}(p_{T})$ from~\cite{pi0AN}. Both the $\pi^{0}$ $A_{N}(p_{T})$ and $\pi^{\pm}$ $A_{N}(p_{T})$ results are consistent with zero within the measured range, but the $\pi^{\pm}$ $A_{N}(p_{T})$ results indicate that $\pi^{+}$ and $\pi^{-}$ may behave differently, potentially exhibiting a dependence on quark flavor.  

\section{Conclusion}
TSSAs provide access to non-perturbative spin-momentum and spin-spin correlations within the proton. While both TMD factorization and twist 3 collinear factorization can explain large measured TSSAs, twist 3 collinear factorization requires only a single scale (the hard scale), for which the measured particle's $p_{T}$ can be taken as a proxy. Measurements presented in this note can therefore be used to constrain twist 3 correlators. Midrapidity heavy flavor $e^{\pm}$ and $\pi^{\pm}$ TSSAs from PHENIX were presented for $p + p^{\uparrow}$ collisions at $\sqrt{s} = 200$ GeV. The heavy flavor $e^{\pm}$ TSSA is the most precise to date, and is consistent with zero in the measured range for both the charge combined and separated samples. The charge separated heavy flavor $e^{+,-}$ TSSAs were compared with theoretical predictions from~\cite{trigluon_twists}, and will serve useful in providing constraints to the trigluon correlator. The $\pi^{\pm}$ TSSAs are also consistent with zero in the measured range, but the different charges seem to display different trends, which may point to flavor sensitivity of the separate charges. The $\pi^{\pm}$ TSSAs were also compared with previously measured $\pi^{0}$ TSSAs from~\cite{pi0AN}. Both presented results are in preparation for publication, and provide additional insight on gluon spin-momentum correlations inside protons
\section*{Acknowledgements}
Thank you to Zhongbo Kang for providing the parametrizations for the trigluon correlator curves from~\cite{trigluon_twists} shown in Figure~\ref{fig:AN_chargesep_theory}.
\paragraph{Funding information}
The funding for this project was provided by the Department of Energy, grant number DE-SC0013393.



\bibliography{SciPost_Example_BiBTeX_File.bib}

\nolinenumbers

\end{document}